\documentclass[conference]{IEEEtran}
\usepackage{cite}
\usepackage{amsmath,amssymb,amsfonts}
\usepackage{algorithmic}
\usepackage{graphicx}
\usepackage{textcomp}
\usepackage{xcolor}
\def\BibTeX{{\rm B\kern-.05em{\sc i\kern-.025em b}\kern-.08em
    T\kern-.1667em\lower.7ex\hbox{E}\kern-.125emX}}

\begin{document}

\title{A proposal for a hybrid free-space optical quantum communication network with hexagonal boron nitride-based single photon sources}

\author{\IEEEauthorblockN{Julien Chénedé, Mostafa Abasifard, Tjorben Matthes, Aslı Çakan, Tobias Vogl}
\IEEEauthorblockA{\textit{Department of Computer Engineering}\\
\textit{TUM School of Computation, Information and Technology}\\
\textit{Technical University of Munich}\\
Munich, Germany\\
\textit{ }
}
\IEEEauthorblockA{\textit{Munich Center for Quantum Science and Technology (MCQST)}\\
Munich, Germany\\
\textit{ }\\
Email: tobias.vogl@tum.de}
}

\maketitle

\begin{abstract}
Hexagonal boron nitride (hBN) is known as a promising solid-state platform to host room temperature quantum emitters that produce high purity single photons. The bright and spectrally adaptable hBN emitters are space-compatible, making hBN well-suited for satellite-based free-space optical (FSO) quantum key distribution (QKD) at wavelengths where the atmospheric background is naturally suppressed. This paper presents a pathway to develop hBN emitters operating near the Ca-II Fraunhofer line (at 854 nm), enabling daylight FSO operation. Moreover, due to its compatibility with the first telecommunication window at 850 nm, it is possible to interface with optical fibers to bridge the `last mile' in a scenario where multiple end-users connect through a single optical ground station to a QKD satellite. We therefore introduce a concept that combines continuous operation of a quantum network, hybrid links to minimize deployment costs, and high data rates due to the use of realistic single photon sources. The realization would be an important milestone for the development of the quantum internet.
\end{abstract}

\begin{IEEEkeywords}
quantum communication, fluorescent defects, hexagonal boron nitride, single photons, hybrid free-space and fiber networks
\end{IEEEkeywords}

\section{Introduction}
The modern internet underpins critical services such as banking and government infrastructure and therefore relies on robust cryptographic key exchange. The anticipated arrival of large-scale quantum computers threatens current asymmetric schemes (e.g., RSA cryptosystem~\cite{RSA78}) via algorithms such as Shor's~\cite{Shor94, Shor97}; consequently, alternatives such as quantum key distribution (QKD) and post-quantum cryptography (PQC)~\cite{Pirandola2020} are being actively pursued. QKD provides information-theoretic secrecy founded on quantum mechanics and is a promising component of a future hybrid security stack. QKD achieves forward and backward secrecy by encoding information in single-photon quantum states whose measurements unavoidably disturb them, preventing undetected eavesdropping. Of course, it is believed that these laws will (i) hold independently of the resources of any eavesdropper, and (ii) will also hold in the future.\\

While PQC can be implemented purely in software, QKD requires new hardware that makes it expensive to deploy at large scales. A future quantum internet~\cite{doi:10.1126/science.aam9288} (the internet as we know it with end-to-end quantum encryptions) might therefore be hybrid, where highly sensitive communication (e.g., governments, intelligence agencies, military, large companies, data centers, etc.) is encrypted with QKD and civil end-user communication (email, instant messages, etc.) is secured with PQC. The hardware needed for QKD includes single photon sources and quantum state modulators on the transmitter side, optical fibers and free-space links as a quantum channel, as well as quantum state analyzers and single photon detectors on the receiver side.\\

In this work, we propose a scheme for a quantum network implementation based on free-space optical (FSO) and fiber-based QKD links. We first develop a model that identifies the ideal wavelength for hybrid quantum communication and describe an interface to a fiber-based quantum network that is currently under construction at the Technical University of Munich (TUM). In addition, we review quantum emitters hosted by hexagonal boron nitride (hBN) as practical single photon sources that can be used in the QKD transmitter to generate the required quantum states.

\section{Free-space optical quantum communication}
\begin{figure*}[htbp]
\centerline{\includegraphics{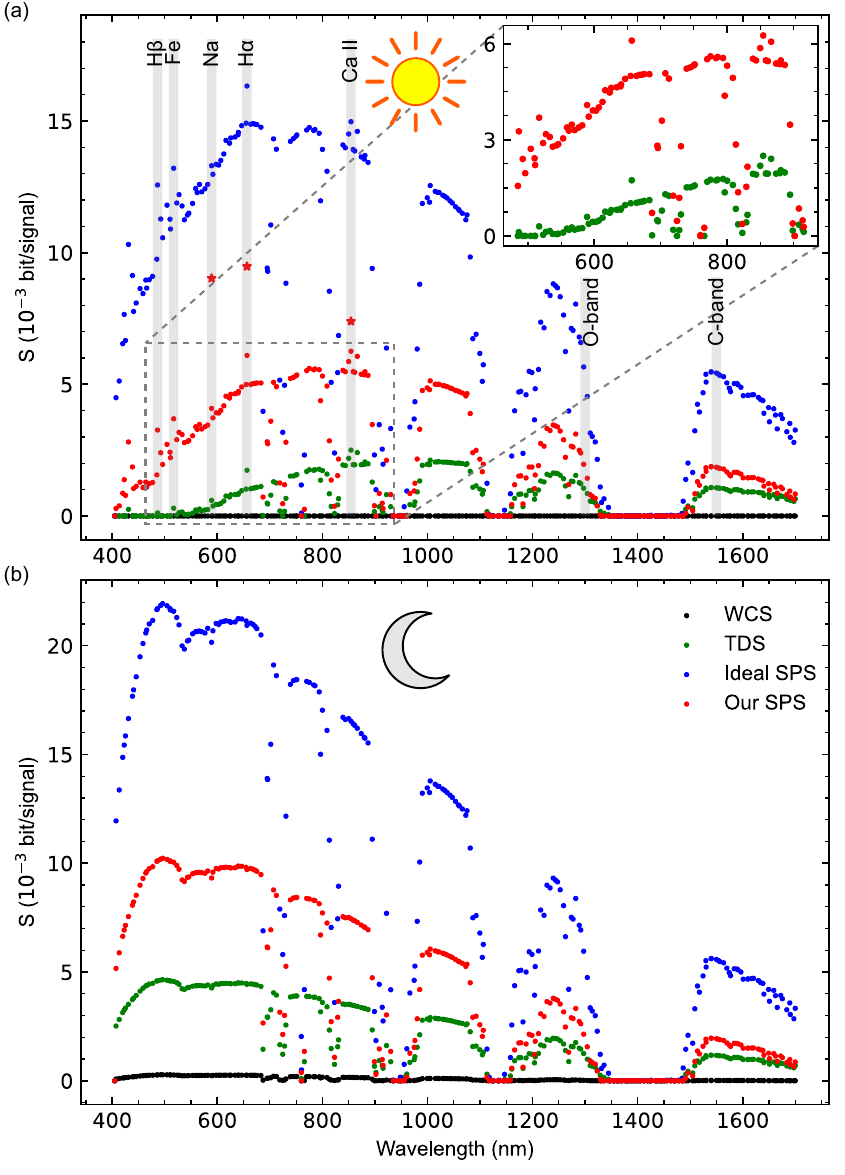}}
\caption{The extractable secret bit per signal is shown for a single satellite (zenith) pass at an altitude of 500 km, evaluated for (a) daylight and (b) nighttime conditions. In daylight, several Fraunhofer lines,  which are indicated by the labels of their corresponding elements, exhibit noticeably improved performance due to reduced background light. The ‘*’ symbols indicate the key rates achievable when using specialized Fraunhofer filters (commercially available for selected lines) that are commonly employed in solar astronomy. At night, where background noise is negligible, no particular wavelength offers a unique advantage; nonetheless, visible wavelengths generally outperform those in the infrared. The results are based on the asymptotic model. Blue/red/green/black data points correspond to an ideal SPS/a realistic SPS that we have developed/two decoy states/weak coherent states, respectively. This figure is adapted from Ref.~\cite{Abasifard2024} with permission from AIP Publishing available under a CC-BY 4.0 license. Copyright 2024 Abasifard \textit{et al.}
}
\label{fig}
\end{figure*}
Perhaps the most important choice for the realization of a quantum communication link is the quantum channel. It is critical, that this channel satisfies the following requirements: it must (i) guide the photons from the sender to the receiver, (ii) maintain the quantum states, (iii) have low loss at the photon wavelength, (iv) prevent any background light from leaking into the channel; while ideally also (v) being continuously available and (vi) have a low technological complexity. In general, there are two options: optical fibers or free-space links. Optical fibers suffer from losses due to molecular absorption and Rayleigh scattering. These loss mechanisms are minimal in the telecom C-band at 1550 nm, with commercial ultra-low-loss fibers achieving attenuation coefficients of around 0.15 dB/km. State-of-the-art fibers can even achieve below 0.1 dB/km, but these are not yet commercially available~\cite{Petrovich2025}. With commercial fibers, national distances over 100s of km can be overcome. Twin-field QKD protocols~\cite{Lucamarini2018} can double this due to a square-root scaling of the transmission efficiency with distance and have set the current distance record of fiber-based QKD at 1002 km~\cite{Liu2023}.\\

Longer distances can be achieved by satellites, as scattering, absorption, and turbulence above 10 km in the atmosphere become negligible~\cite{RevModPhys.94.035001}. The losses are then dominated by diffraction, which in turn depends on the telescope sizes. QKD with satellites has been demonstrated with the Micius~\cite{Liao2017} and Jinan-1 satellites~\cite{Li2025}. A satellite link can only be established when there is a direct line of sight between an optical ground station (OGS) and the satellite. Intercontinental connections can still be realized when the satellite serves as a trusted node: a key is first exchanged with OGS 1, and as the satellite orbits around the Earth later with OGS 2. The satellite can broadcast the XOR value of both keys (bit-wise sum modulo 2), therefore perfectly mixing two random keys. Both OGSs can simply apply the XOR to their own key and the broadcasted key and will obtain the other OGS's key (their own key will be bit-wise added twice modulo 2). This has been demonstrated using Micius between Austria and China over 7600 km~\cite{PhysRevLett.120.030501}.\\

A major limitation of free-space links, however, is that during the day, sunlight saturates sensitive single photon detectors, making it impossible to resolve the much weaker quantum signal (single photons). This violates requirements (iv) and (v) and significantly limits application scenarios. Moreover, it drives up the cost of a QKD satellite constellation (the constellation would be idle during the day and during the night; more satellites are needed to cope with the increased demand to make up for the idle time). This issue can be resolved by QKD systems that are daylight compatible~\cite{Ko2018, Gong2018, Avesani2021}. One option for this is using longer wavelengths in the near-infrared (NIR), where the solar background is intrinsically weaker. This, however, comes at the expense that diffraction is proportional to the wavelength, i.e., the telescope sizes must be larger, which further drives cost (the price of a telescope scales exponentially with the diameter), as well as commercial silicon-based single photon detectors cannot be used anymore. Detectors based on, e.g., InGaAs are typically less efficient, and highly efficient superconducting nanowire single photon detectors require cryostats~\cite{Natarajan_2012}, which again increases the cost of ground station equipment.\\

An alternative using visible wavelengths was proposed originally in 2006~\cite{Rogers2006}: when wavelengths at the Fraunhofer lines in the solar spectrum are used, one can filter around these lines in the receiver and separate single photons from the strong solar radiance. For some Fraunhofer lines, such as H$\alpha$ at 656 nm and Ca-II at 854 nm, there are narrow-band filter cavities commercially available, as they are being used in solar observation and astronomy. The question arises, which Fraunhofer line yields the best performance, i.e., the highest data rate. This is usually quantified in terms of secret key rate, which is the final rate of a QKD system that can be used for encryption of the actual data after error correction and privacy amplification.\\

We have modelled space-to-ground links, taking into account atmospheric transmission, diffraction losses, pointing errors, the solar background spectrum, as well as state-of-the-art performance characteristics of single photon detectors and telescope sizes of the Micius mission to spectrally resolve the achievable secret key rate (in units of secret bit per signal)~\cite{Abasifard2024}. This allows us to select the ideal wavelength for free-space daylight QKD that can be achieved in a realistic scenario with the technology available today. Moreover, we compared the performance for different photon source types: an ideal single photon source (SPS), a realistic SPS based on a defect in hBN that we have fabricated~\cite{Vogl:2019-QC}, weak coherent states (WCS) from a laser, and two decoy states (TDS, which are WCS with a varying intensity)~\cite{Abasifard2024}. The results are shown in Figure \ref{fig}. The annual integration of the secret key rate accounts for temporal variations (e.g., seasonal and diurnal atmospheric changes) and spatial effects (e.g., elevation-dependent transmission and non-zenith passes), providing a robust yearly average performance. The instrument parameters uncertainties in the secret key rate calculations are based on manufacturer datasheets for detectors and the pointing accuracy from the Micius mission, as detailed in Ref\cite{Abasifard2024}. We observe two general trends: visible wavelengths perform better than NIR wavelengths, and realistic SPSs are already sufficient to outperform laser-based QKD. As expected, the wavelengths of the Fraunhofer lines peak due to the reduced background light at those wavelengths. The H$\alpha$ wavelength at 656 nm is the overall best wavelength in daylight conditions (see Figure \ref{fig}(a)). During the night (see Figure \ref{fig}(b)), there is no advantage of the Fraunhofer lines as there is no solar background, resulting in a smoother spectrum. It should be noted that the night model assumed no background at all, similar to new moon conditions. For a full moon, this might not be satisfied; however, the Fraunhofer lines are also visible in the moonlight (as it is just scattered sunlight), but due to inelastic scattering, the lines are typically broader and less deep. The key rates during the night are also overall higher compared to during the day, due to the finite filtering bandwidth of the solar background suppression. Another good wavelength is the Ca-II line at 854 nm. This wavelength has the appeal that it is near 850 nm, the standard wavelength of free-space QKD~\cite{Abasifard2024}. This choice makes daylight QKD directly compatible with basically any other FSO link.

\section{Interface to fiber networks}
FSO quantum communication with a satellite constellation described above makes it possible to establish a global (trusted node) quantum network. While the satellites are undoubtedly the most expensive hardware components of the network, the OGS equipment must not be underestimated. It is hardly imaginable that every end-user of the quantum network will operate their own ground station (e.g., on their roof). It is more likely that there is one or a few OGSs within or near a metropolitan region, and the `last mile' to the end-users is bridged by optical fibers. This is incompatible with the H$\alpha$ wavelength. Optical fibers have typical attenuation coefficients exceeding 10 dB/km at 630 nm (e.g., S630-HP), making the last few km very inefficient. To maintain the free-space wavelength advantages in daylight conditions, the first telecom window at 850 nm is compatible, with attenuation coefficients in optical fibers down to 2 dB/km. For scenarios requiring fiber connectivity, 854 nm delivers 72\% of the annual secret key length compared to 656 nm. Hence, the 854 nm wavelength is selected for its compatibility with fiber-optic networks, balancing free-space performance and low fiber attenuation to enable practical end-user connectivity. \\

Such a hybrid channel link from a satellite into a fiber network without the OGS serving as a trusted node has not yet been demonstrated. It should be noted that making the satellite a trusted node is a security compromise that can be acceptable, as a satellite cannot be accessed practically in space (the only demonstration of this was the servicing of the Hubble telescope), and hacking can be effectively prevented by accepting only encrypted communication. An OGS, in turn, could be accessed in principle unnoticed.\\

With the Quantum Network at the Technical University of Munich (QuaNTUM), we are currently deploying a test infrastructure that demonstrates the feasibility of the concept of an OGS relay station. QuaNTUM is under construction on TUM's high-tech research campus in Garching (near Munich) and will link all research institutes on campus that investigate quantum technologies. In the first phase, the fibers will be deployed, and quantum communication demonstrated. This is not a trusted node network as has been demonstrated in Boston~\cite{elliott2004darpaquantumnetwork}, Vienna~\cite{Peev_2009}, Tokyo~\cite{Sasaki:11}, and China~\cite{Chen2021}; rather the network will enable full quantum communication (including protocols like quantum teleportation, etc.), over end-to-end polarization-compensated and time-synchronized \cite{Häusler2023} fiber links with available wavelengths ranging from 780 to 1600 nm. It is therefore possible to transfer quantum information from, e.g., trapped ions (such as Rb) to quantum dots. A polarization control device is under development to enable active closed-loop compensation in the QuaNTUM network. In the next phase, our central node of the star-shaped QuaNTUM will be equipped with an OGS.

\section{Single photon sources}
We finally turn to the question of which single photon source can be used. We use the fluorescence of defects in the 2D material hexagonal boron nitride~\cite{Tran:2016a, Tran:2016b}. These quantum emitters are known to feature a high single photon luminosity~\cite{Tran:2016a}, which is due to high internal quantum efficiencies~\cite{Nikolay19}, short excited state lifetimes~\cite{Vogl2018-do}, and near-ideal photon extraction. The latter is because the emitters typically have an in-plane dipole, resulting in a strong out-of-plane emission component. Moreover, in a (few-layer) 2D crystal, the defects are not surrounded by any high refractive index material, which restricts the extraction efficiency due to total internal and Fresnel reflection at the crystal-air interface\cite{Vogl2019-ns}. The hBN emitters can operate in space environments~\cite{10.1038/s41467-019-09219-5} and over a large temperature range from 4 up to 800 K~\cite{Kianinia:2017} and are robust to aging~\cite{Vogl2018-do}.\\

To implement the daylight-compatible QKD scenario in hybrid FSO and fiber channels, a wavelength of 854 nm is required. Hexagonal boron nitride is known to host a large number of defects, with their optical emission wavelengths ranging from the UV up to NIR (based on theoretical simulations~\cite{Cholsuk2024a}). Experimentally, emitters in the UV~\cite{Bourrellier:2016}, entire visible spectrum\cite{Tran:2016b,Dietrich2018,https://doi.org/10.1002/adom.202402508}, and up to 1000 nm~\cite{10.1063/5.0008242} have been demonstrated. The negatively-charged boron vacancy (V$^-_\textrm{B}$) has a spectral overlap with 854 nm~\cite{doi:10.1021/acs.nanolett.2c00739} and can be fabricated using ion irradiation (with noble gas ions~\cite{PhysRevMaterials.9.056203}) or high energy electrons~\cite{https://doi.org/10.1002/smll.202301926}. The knock-on damage threshold energy to displace a boron atom requires a high particle acceleration voltage, but then typically many vacancies are created, resulting in defect ensembles and it is difficult to isolate single defects~\cite{https://doi.org/10.1002/smll.202301926}. Our approach is to use an extrinsic defect and implant an impurity, or activate a defect (that is already there) with electron irradiation.\\

Our theoretical simulations~\cite{Cholsuk2024a} predict that the following defects have their peak emission near 854 nm:
\begin{itemize}
    \item Ga$_\textrm{B}$V$_\textrm{N}$ at 831 nm
    \item Al$_\textrm{N}$ at 834 nm
    \item Ga$_\textrm{N}$ at 845 nm
    \item O$_\textrm{B}$V$_\textrm{N}$ at 881 nm
\end{itemize}

It is known that the precise emission wavelength can be fine-tuned by crystal strain~\cite{Grosso2017,nano12142427}, which works well in 2D materials. An alternative approach that we have developed is near-deterministically activating emitters using a low-energy electron irradiation~\cite{Kumar:2023}. This technique allows us already to fabricate green-yellow emitters with high yield and high photon purity~\cite{doi:10.1021/acsnano.3c08940} that would be compatible with the Na-D Fraunhofer lines, but not with the Ca-II line at 854 nm. A promising pathway to fabricate the needed emitters could be to combine both methods: large area ion irradiation and localized activation with the electron beam.

\section{Conclusion}
To summarize, in this work, we have outlined a scheme to implement satellite-based FSO quantum communication that can operate in daylight conditions and is compatible with fiber networks. The dark Fraunhofer lines in the solar spectrum are exploited to enable the separation and suppression of sunlight. This will allow using visible (shorter) wavelengths, which are less affected by diffraction and therefore eliminate the need for large telescopes. Moreover, cheap silicon single photon detectors that are commercially available can be used to reduce the deployment cost in the end-user devices (i.e., mass production is possible). With only a small secret key rate compromise, the Ca-II Fraunhofer line at 854 nm can be used, which coincides with the first telecommunication window. In an application scenario, this allows one to bridge the `last mile' between the optical ground station and an end-user in a metropolitan fiber network. As a quantum light source, we propose defects in hBN that cover a large wavelength range, but need to be synthesized at 854 nm.\\

With the expected arrival of large-scale quantum computers, it is essential that the current asymmetric cryptography infrastructure is replaced quickly. As hardware developments and subsequent deployments, as well as standardization, require time, it is critical to start the transition now. The proposed scheme improves the practicality of QKD and, therefore, can increase the adoption rate. We believe that it will find many applications in the development of the quantum internet.

\section*{Acknowledgment}
This research is part of the Munich Quantum Valley, which is supported by the Bavarian state government with funds from the Hightech Agenda Bayern Plus. This work was funded by the Deutsche Forschungsgemeinschaft (DFG, German Research Foundation) under Germany’s Excellence Strategy- EXC-2111-390814868 (MCQST). The authors acknowledge support from the Federal Ministry of Research, Technology and Space (BMFTR) under grant 13N16292 (ATOMIQS).

\bibliographystyle{IEEEtran}
\bibliography{IEEEexample}

@PREAMBLE{
 "\providecommand{\noopsort}[1]{}" 
 # "\providecommand{\singleletter}[1]{#1}%" 
}

@article{Liu2023,
  title = {Experimental Twin-Field Quantum Key Distribution over 1000 km Fiber Distance},
  author = {Liu, Yang and Zhang, Wei-Jun and Jiang, Cong and Chen, Jiu-Peng and Zhang, Chi and Pan, Wen-Xin and Ma, Di and Dong, Hao and Xiong, Jia-Min and Zhang, Cheng-Jun and Li, Hao and Wang, Rui-Chun and Wu, Jun and Chen, Teng-Yun and You, Lixing and Wang, Xiang-Bin and Zhang, Qiang and Pan, Jian-Wei},
  journal = {Phys. Rev. Lett.},
  volume = {130},
  issue = {21},
  pages = {210801},
  numpages = {6},
  year = {2023},
  publisher = {American Physical Society},
  doi = {10.1103/PhysRevLett.130.210801},
  url = {https://link.aps.org/doi/10.1103/PhysRevLett.130.210801}
}

@article{RSA78,
author = {Rivest, R. L. and Shamir, A. and Adleman, L.},
title = {A method for obtaining digital signatures and public-key cryptosystems},
year = {1978},
issue_date = {Feb. 1978},
publisher = {Association for Computing Machinery},
address = {New York, NY, USA},
volume = {21},
number = {2},
issn = {0001-0782},
journal = {Commun. ACM},
pages = {120–126},
numpages = {7},
url={https://doi.org/10.1145/359340.359342},
}

@inproceedings{Rogers2006,
author = {D. J. Rogers and J. C. Bienfang and A. Mink and B. J. Hershman and A. Nakassis and X. Tang and L. Ma and D. H. Su and Carl J. Williams and Charles W. Clark},
title = {{Free-space quantum cryptography in the H-alpha Fraunhofer window}},
volume = {6304},
booktitle = {Free-Space Laser Communications VI},
editor = {Arun K. Majumdar and Christopher C. Davis},
organization = {International Society for Optics and Photonics},
publisher = {SPIE},
pages = {630417},
year = {2006},
url={https://doi.org/10.1117/12.680899},
}

@INPROCEEDINGS{Shor94,
  author={Shor, P.W.},
  booktitle={Proceedings 35th Annual Symposium on Foundations of Computer Science}, 
  title={Algorithms for quantum computation: discrete logarithms and factoring}, 
  year={1994},
  volume={},
  number={},
  pages={124-134},
  doi={10.1109/SFCS.1994.365700},
url={https://doi.org/10.1109/SFCS.1994.365700},
}

@article{Shor97,
author = {Shor, Peter W.},
title = {Polynomial-Time Algorithms for Prime Factorization and Discrete Logarithms on a Quantum Computer},
journal = {SIAM Journal on Computing},
volume = {26},
number = {5},
pages = {1484-1509},
year = {1997},
doi = {10.1137/S0097539795293172},
url={https://doi.org/10.1137/S0097539795293172},
}

@article{Pirandola2020,
author = {S. Pirandola and U. L. Andersen and L. Banchi and M. Berta and D. Bunandar and R. Colbeck and D. Englund and T. Gehring and C. Lupo and C. Ottaviani and J. L. Pereira and M. Razavi and J. Shamsul Shaari and M. Tomamichel and V. C. Usenko and G. Vallone and P. Villoresi and P. Wallden},
journal = {Adv. Opt. Photon.},
number = {4},
pages = {1012--1236},
publisher = {Optica Publishing Group},
title = {Advances in quantum cryptography},
volume = {12},
year = {2020},
url={https://doi.org/10.1364/AOP.361502},
}

@article{Li2025,
  author    = {Yang Li and Wen-Qi Cai and Ji-Gang Ren and Chao-Ze Wang and Meng Yang and Liang Zhang and Hui-Ying Wu and Liang Chang and Jin-Cai Wu and Biao Jin and Hua-Jian Xue and Xue-Jiao Li and Hui Liu and Guang-Wen Yu and Xue-Ying Tao and ...},
  title     = {Microsatellite-based real-time quantum key distribution},
  journal   = {Nature},
  year      = {2025},
  volume    = {640},
  number    = {8057},
  pages     = {47--54},
  issn      = {1476-4687},
  doi       = {10.1038/s41586-025-08739-z},
  url       = {https://doi.org/10.1038/s41586-025-08739-z}
}

@article{Gong2018,
author = {Yun-Hong Gong and Kui-Xing Yang and Hai-Lin Yong and Jian-Yu Guan and Guo-Liang Shentu and Chang Liu and Feng-Zhi Li and Yuan Cao and Juan Yin and Sheng-Kai Liao and Ji-Gang Ren and Qiang Zhang and Cheng-Zhi Peng and Jian-Wei Pan},
journal = {Opt. Express},
number = {15},
pages = {18897--18905},
publisher = {Optica Publishing Group},
title = {Free-space quantum key distribution in urban daylight with the SPGD algorithm control of a deformable mirror},
volume = {26},
year = {2018},
url = {https://opg.optica.org/oe/abstract.cfm?URI=oe-26-15-18897},
doi = {10.1364/OE.26.018897},
}

@article{Liao2017,
  author    = {Sheng-Kai Liao and Hai-Lin Yong and Chang Liu and Guo-Liang Shentu and Dong-Dong Li and Jin Lin and Hui Dai and Shuang-Qiang Zhao and Bo Li and Jian-Yu Guan and Wei Chen and Yun-Hong Gong and Yang Li and Ze-Hong Lin and ... and Paolo Villoresi},
  title     = {Satellite-to-ground quantum key distribution},
  journal   = {Nature Photonics},
  year      = {2017},
  volume    = {11},
  pages     = {509--513},
  doi       = {10.1038/nphoton.2017.116},
  url       = {https://doi.org/10.1038/nphoton.2017.116}
}

@article{Avesani2021,
  author    = {M. Avesani and L. Calderaro and M. Schiavon and A. Stanco and C. Agnesi and A. Santamato and M. Zahidy and A. Scriminich and G. Foletto and G. Contestabile and ...},
  title     = {Full daylight quantum-key-distribution at 1550 nm enabled by integrated silicon photonics},
  journal   = {npj Quantum Information},
  year      = {2021},
  volume    = {7},
  number    = {1},
  pages     = {93},
  doi       = {10.1038/s41534-021-00421-2},
  url       = {https://doi.org/10.1038/s41534-021-00421-2}
}

@ARTICLE{Vogl2018-do,
author = {Vogl, Tobias and Campbell, Geoff and Buchler, Ben C. and Lu, Yuerui and Lam, Ping Koy},
title = {Fabrication and Deterministic Transfer of High-Quality Quantum Emitters in Hexagonal Boron Nitride},
journal = {ACS Photonics},
volume = {5},
number = {6},
pages = {2305-2312},
year = {2018},
doi = {10.1021/acsphotonics.8b00127},
URL = {https://doi.org/10.1021/acsphotonics.8b00127},
eprint = {https://doi.org/10.1021/acsphotonics.8b00127}
}

@article{Nikolay19,
author = {Niko Nikolay and Noah Mendelson and Ersan \"{O}zelci and Bernd Sontheimer and Florian B\"{o}hm and G\"{u}nter Kewes and Milos Toth and Igor Aharonovich and Oliver Benson},
journal = {Optica},
number = {8},
pages = {1084--1088},
publisher = {Optica Publishing Group},
title = {Direct measurement of quantum efficiency of single-photon emitters in hexagonal boron nitride},
volume = {6},
year = {2019},
url = {https://opg.optica.org/optica/abstract.cfm?URI=optica-6-8-1084},
doi = {10.1364/OPTICA.6.001084},
}

@article{Vogl2019-ns,
   title={Atomic localization of quantum emitters in multilayer hexagonal boron nitride},
   volume={11},
   ISSN={2040-3372},
   url={http://dx.doi.org/10.1039/C9NR04269E},
   DOI={10.1039/c9nr04269e},
   number={30},
   journal={Nanoscale},
   publisher={Royal Society of Chemistry (RSC)},
   author={Vogl, Tobias and Doherty, Marcus W. and Buchler, Ben C. and Lu, Yuerui and Lam, Ping Koy},
   year={2019},
   pages={14362–14371} }

@article{Dietrich2018,
  title = {Observation of Fourier transform limited lines in hexagonal boron nitride},
  author = {Dietrich, A. and B\"urk, M. and Steiger, E. S. and Antoniuk, L. and Tran, T. T. and Nguyen, M. and Aharonovich, I. and Jelezko, F. and Kubanek, A.},
  journal = {Phys. Rev. B},
  volume = {98},
  issue = {8},
  pages = {081414},
  numpages = {5},
  year = {2018},
  publisher = {American Physical Society},
  doi = {10.1103/PhysRevB.98.081414},
  url = {https://link.aps.org/doi/10.1103/PhysRevB.98.081414}
}

@article{10.1063/5.0008242,
    author = {Camphausen, Robin and Marini, Loris and Tawfik, Sherif Abdulkader and Tran, Toan Trong and Ford, Michael J. and Palomba, Stefano},
    title = {Observation of near-infrared sub-Poissonian photon emission in hexagonal boron nitride at room temperature},
    journal = {APL Photonics},
    volume = {5},
    number = {7},
    pages = {076103},
    year = {2020},
    issn = {2378-0967},
    doi = {10.1063/5.0008242},
    url = {https://doi.org/10.1063/5.0008242},
    eprint = {https://pubs.aip.org/aip/app/article-pdf/doi/10.1063/5.0008242/13386733/076103_1_online.pdf},
}

@article{https://doi.org/10.1002/adom.202402508,
author = {Çakan, Aslı and Cholsuk, Chanaprom and Gale, Angus and Kianinia, Mehran and Paçal, Serkan and Ateş, Serkan and Aharonovich, Igor and Toth, Milos and Vogl, Tobias},
title = {Quantum Optics Applications of Hexagonal Boron Nitride Defects},
journal = {Advanced Optical Materials},
volume = {13},
number = {7},
pages = {2402508},
doi = {https://doi.org/10.1002/adom.202402508},
url = {https://advanced.onlinelibrary.wiley.com/doi/abs/10.1002/adom.202402508},
eprint = {https://advanced.onlinelibrary.wiley.com/doi/pdf/10.1002/adom.202402508},
year = {2025}
}

@Article{nano12142427,
AUTHOR = {Cholsuk, Chanaprom and Suwanna, Sujin and Vogl, Tobias},
TITLE = {Tailoring the Emission Wavelength of Color Centers in Hexagonal Boron Nitride for Quantum Applications},
JOURNAL = {Nanomaterials},
VOLUME = {12},
YEAR = {2022},
NUMBER = {14},
ARTICLE-NUMBER = {2427},
URL = {https://www.mdpi.com/2079-4991/12/14/2427},
}

@Article{Grosso2017,
author={Grosso, Gabriele
and Moon, Hyowon
and Lienhard, Benjamin
and Ali, Sajid
and Efetov, Dmitri K.
and Furchi, Marco M.
and Jarillo-Herrero, Pablo
and Ford, Michael J.
and Aharonovich, Igor
and Englund, Dirk},
title={Tunable and high-purity room temperature single-photon emission from atomic defects in hexagonal boron nitride},
journal={Nature Communications},
year={2017},
day={26},
volume={8},
number={1},
pages={705},
doi={10.1038/s41467-017-00810-2},
url={https://doi.org/10.1038/s41467-017-00810-2}
}

@article{Chen2021, 
year = {2021}, 
title = {{An integrated space-to-ground quantum communication network over 4,600 kilometres}}, 
author = {Chen, Yu-Ao and Zhang, Qiang and Chen, Teng-Yun and Cai, Wen-Qi and Liao, Sheng-Kai and Zhang, Jun and Chen, Kai and Yin, Juan and Ren, Ji-Gang and Chen, Zhu and Han, Sheng-Long and Yu, Qing and Liang, Ken and Zhou, Fei and Yuan, Xiao and Zhao, Mei-Sheng and Wang, Tian-Yin and Jiang, Xiao and Zhang, Liang and Liu, Wei-Yue and Li, Yang and Shen, Qi and Cao, Yuan and Lu, Chao-Yang and Shu, Rong and Wang, Jian-Yu and Li, Li and Liu, Nai-Le and Xu, Feihu and Wang, Xiang-Bin and Peng, Cheng-Zhi and Pan, Jian-Wei}, 
journal = {Nature}, 
issn = {0028-0836}, 
doi = {10.1038/s41586-020-03093-8}, 
pmid = {33408416}, 
pages = {214--219}, 
number = {7841}, 
volume = {589},
url = {https://doi.org/10.1038/s41586-020-03093-8}, 
}

@article{Lucamarini2018, 
year = {2018}, 
title = {{Overcoming the rate–distance limit of quantum key distribution without quantum repeaters}}, 
author = {Lucamarini, M. and Yuan, Z. L. and Dynes, J. F. and Shields, A. J.}, 
journal = {Nature}, 
issn = {0028-0836}, 
doi = {10.1038/s41586-018-0066-6}, 
pmid = {29720656}, 
pages = {400--403}, 
number = {7705}, 
volume = {557}, 
url = {http://dx.doi.org/10.1038/s41586-018-0066-6}, 
}

@article{Ko2018, 
year = {2018}, 
title = {{Experimental filtering effect on the daylight operation of a free-space quantum key distribution}}, 
author = {Ko, Heasin and Kim, Kap-Joong and Choe, Joong-Seon and Choi, Byung-Seok and Kim, Jong-Hoi and Baek, Yongsoon and Youn, Chun Ju}, 
journal = {Scientific Reports}, 
doi = {10.1038/s41598-018-33699-y}, 
pmid = {30333620}, 
pmcid = {PMC6193016},
pages = {15315}, 
number = {1}, 
volume = {8},
url = {http://dx.doi.org/10.1038/s41598-018-33699-y}, 
}

@article{Abasifard2024,
    author = {Abasifard, Mostafa and Cholsuk, Chanaprom and Pousa, Roberto G. and Kumar, Anand and Zand, Ashkan and Riel, Thomas and Oi, Daniel K. L. and Vogl, Tobias},
    title = "{The ideal wavelength for daylight free-space quantum key distribution}",
    journal = {APL Quantum},
    volume = {1},
    number = {1},
    pages = {016113},
    year = {2024},
    issn = {2835-0103},
    doi = {10.1063/5.0186767},
    url = {https://doi.org/10.1063/5.0186767},
}

@article{Cholsuk2024a,
author = {Cholsuk, Chanaprom and Zand, Ashkan and Çakan, Aslı and Vogl, Tobias},
title = {The hBN Defects Database: A Theoretical Compilation of Color Centers in Hexagonal Boron Nitride},
journal = {The Journal of Physical Chemistry C},
volume = {128},
number = {30},
pages = {12716-12725},
year = {2024},
doi = {10.1021/acs.jpcc.4c03404},
URL = {https://doi.org/10.1021/acs.jpcc.4c03404}
}

@article{Bourrellier:2016,
   author = {Bourrellier, Romain and Meuret, Sophie and Tararan, Anna and Stéphan, Odile and Kociak, Mathieu and Tizei, Luiz HG and Zobelli, Alberto},
   title = {Bright UV single photon emission at point defects in h-BN},
   doi = {10.1021/acs.nanolett.6b01368},
   journal = {Nano letters},
   volume = {16},
   number = {7},
   pages = {4317-4321},
   ISSN = {1530-6984},
   year = {2016},
   url = {https://doi.org/10.1021/acs.nanolett.6b01368},
}

@article{Kumar:2023,
   author = {Kumar, Anand and Cholsuk, Chanaprom and Zand, Ashkan and Mishuk, Mohammad N. and Matthes, Tjorben and Eilenberger, Falk and Suwanna, Sujin and Vogl, Tobias},
   title = {Localized creation of yellow single photon emitting carbon complexes in hexagonal boron nitride},
   journal = {APL Materials},
   volume = {11},
   number = {7},
   pages = {071108},
   ISSN = {2166-532X},
   DOI = {10.1063/5.0147560},
   url = {https://doi.org/10.1063/5.0147560},
   year = {2023},
   type = {Journal Article}
}

@article{doi:10.1126/science.aam9288,
author = {Stephanie Wehner  and David Elkouss  and Ronald Hanson },
title = {Quantum internet: A vision for the road ahead},
journal = {Science},
volume = {362},
number = {6412},
pages = {eaam9288},
year = {2018},
doi = {10.1126/science.aam9288},
URL = {https://www.science.org/doi/abs/10.1126/science.aam9288},
eprint = {https://www.science.org/doi/pdf/10.1126/science.aam9288},
}

@Article{Petrovich2025,
author={Petrovich, Marco
and Numkam Fokoua, Eric
and Chen, Yong
and Sakr, Hesham
and Adamu, Abubakar Isa
and Hassan, Rosdi
and Wu, Dong
and Fatobene Ando, Ron
and Papadimopoulos, Athanasios
and Sandoghchi, Seyed Reza
and Jasion, Gregory
and Poletti, Francesco},
title={Broadband optical fibre with an attenuation lower than 0.1 decibel per kilometre},
journal={Nature Photonics},
year={2025},
volume={19},
number={11},
pages={1203-1208},
issn={1749-4893},
doi={10.1038/s41566-025-01747-5},
url={https://doi.org/10.1038/s41566-025-01747-5},
}

@article{RevModPhys.94.035001,
  title = {Micius quantum experiments in space},
  author = {Lu, Chao-Yang and Cao, Yuan and Peng, Cheng-Zhi and Pan, Jian-Wei},
  journal = {Rev. Mod. Phys.},
  volume = {94},
  issue = {3},
  pages = {035001},
  numpages = {46},
  year = {2022},
  doi = {10.1103/RevModPhys.94.035001},
  url = {https://link.aps.org/doi/10.1103/RevModPhys.94.035001}
}

@article{Tran:2016a,
   author = {Tran, Toan Trong and Bray, Kerem and Ford, Michael J. and Toth, Milos and Aharonovich, Igor},
   title = {Quantum emission from hexagonal boron nitride monolayers},
   journal = {Nature Nanotechnology},
   volume = {11},
   number = {1},
   pages = {37-41},
   ISSN = {1748-3395},
   DOI = {10.1038/nnano.2015.242},
   url = {https://doi.org/10.1038/nnano.2015.242},
   year = {2016},
   type = {Journal Article}
}

@article{Tran:2016b,
   author = {Tran, Toan Trong and Elbadawi, Christopher and Totonjian, Daniel and Lobo, Charlene J. and Grosso, Gabriele and Moon, Hyowon and Englund, Dirk R. and Ford, Michael J. and Aharonovich, Igor and Toth, Milos},
   title = {Robust Multicolor Single Photon Emission from Point Defects in Hexagonal Boron Nitride},
   journal = {ACS Nano},
   volume = {10},
   number = {8},
   pages = {7331-7338},
   ISSN = {1936-0851},
   DOI = {10.1021/acsnano.6b03602},
   url = {https://doi.org/10.1021/acsnano.6b03602},
   year = {2016},
   type = {Journal Article}
}

@article{Kianinia:2017,
author = {Kianinia, Mehran and Regan, Blake and Tawfik, Sherif Abdulkader and Tran, Toan Trong and Ford, Michael J. and Aharonovich, Igor and Toth, Milos},
title = {Robust Solid-State Quantum System Operating at 800 K},
journal = {ACS Photonics},
volume = {4},
number = {4},
pages = {768-773},
year = {2017},
doi = {10.1021/acsphotonics.7b00086},
url = {https://doi.org/10.1021/acsphotonics.7b00086}
}

@article{Natarajan_2012,
doi = {10.1088/0953-2048/25/6/063001},
url = {https://doi.org/10.1088/0953-2048/25/6/063001},
year = {2012},
volume = {25},
number = {6},
pages = {063001},
author = {Natarajan, Chandra M and Tanner, Michael G and Hadfield, Robert H},
title = {Superconducting nanowire single-photon detectors: physics and applications},
journal = {Superconductor Science and Technology},
}

@article{PhysRevLett.120.030501,
  title = {Satellite-Relayed Intercontinental Quantum Network},
  author = {Liao, Sheng-Kai and Cai, Wen-Qi and Handsteiner, Johannes and Liu, Bo and Yin, Juan and Zhang, Liang and Rauch, Dominik and Fink, Matthias and Ren, Ji-Gang and Liu, Wei-Yue and Li, Yang and Shen, Qi and Cao, Yuan and Li, Feng-Zhi and Wang, Jian-Feng and Huang, Yong-Mei and Deng, Lei and Xi, Tao and Ma, Lu and Hu, Tai and Li, Li and Liu, Nai-Le and Koidl, Franz and Wang, Peiyuan and Chen, Yu-Ao and Wang, Xiang-Bin and Steindorfer, Michael and Kirchner, Georg and Lu, Chao-Yang and Shu, Rong and Ursin, Rupert and Scheidl, Thomas and Peng, Cheng-Zhi and Wang, Jian-Yu and Zeilinger, Anton and Pan, Jian-Wei},
  journal = {Phys. Rev. Lett.},
  volume = {120},
  issue = {3},
  pages = {030501},
  numpages = {4},
  year = {2018},
  publisher = {American Physical Society},
  doi = {10.1103/PhysRevLett.120.030501},
  url = {https://link.aps.org/doi/10.1103/PhysRevLett.120.030501}
}

@article{Peev_2009,
doi = {10.1088/1367-2630/11/7/075001},
url = {https://doi.org/10.1088/1367-2630/11/7/075001},
year = {2009},
publisher = {},
volume = {11},
number = {7},
pages = {075001},
author = {Peev, M and Pacher, C and Alléaume, R and Barreiro, C and Bouda, J and Boxleitner, W and Debuisschert, T and Diamanti, E and Dianati, M and Dynes, J F and Fasel, S and Fossier, S and Fürst, M and Gautier, J-D and Gay, O and Gisin, N and Grangier, P and Happe, A and Hasani, Y and Hentschel, M and Hübel, H and Humer, G and Länger, T and Legré, M and Lieger, R and Lodewyck, J and Lorünser, T and Lütkenhaus, N and Marhold, A and Matyus, T and Maurhart, O and Monat, L and Nauerth, S and Page, J-B and Poppe, A and Querasser, E and Ribordy, G and Robyr, S and Salvail, L and Sharpe, A W and Shields, A J and Stucki, D and Suda, M and Tamas, C and Themel, T and Thew, R T and Thoma, Y and Treiber, A and Trinkler, P and Tualle-Brouri, R and Vannel, F and Walenta, N and Weier, H and Weinfurter, H and Wimberger, I and Yuan, Z L and Zbinden, H and Zeilinger, A},
title = {The SECOQC quantum key distribution network in Vienna},
journal = {New Journal of Physics},
}

@article{Sasaki:11,
author = {M. Sasaki and M. Fujiwara and H. Ishizuka and W. Klaus and K. Wakui and M. Takeoka and S. Miki and T. Yamashita and Z. Wang and A. Tanaka and K. Yoshino and Y. Nambu and S. Takahashi and A. Tajima and A. Tomita and T. Domeki and T. Hasegawa and Y. Sakai and H. Kobayashi and T. Asai and K. Shimizu and T. Tokura and T. Tsurumaru and M. Matsui and T. Honjo and K. Tamaki and H. Takesue and Y. Tokura and J. F. Dynes and A. R. Dixon and A. W. Sharpe and Z. L. Yuan and A. J. Shields and S. Uchikoga and M. Legr\'{e} and S. Robyr and P. Trinkler and L. Monat and J.-B. Page and G. Ribordy and A. Poppe and A. Allacher and O. Maurhart and T. L\"{a}nger and M. Peev and A. Zeilinger},
journal = {Opt. Express},
number = {11},
pages = {10387--10409},
title = {Field test of quantum key distribution in the Tokyo QKD Network},
volume = {19},
year = {2011},
url = {https://opg.optica.org/oe/abstract.cfm?URI=oe-19-11-10387},
doi = {10.1364/OE.19.010387},
}

@misc{elliott2004darpaquantumnetwork,
      title={The DARPA Quantum Network}, 
      author={Chip Elliott},
      year={2004},
      eprint={quant-ph/0412029},
      archivePrefix={arXiv},
      primaryClass={quant-ph},
      url={https://arxiv.org/abs/quant-ph/0412029}, 
}

@article{PhysRevMaterials.9.056203,
  title = {Quantifying the creation of negatively charged boron vacancies in He-ion irradiated hexagonal boron nitride},
  author = {Carbone, Amedeo and Breev, Ilia D. and Figueiredo, Johannes and Kretschmer, Silvan and Geilen, Leonard and Ben Mhenni, Amine and Arceri, Johannes and Krasheninnikov, Arkady V. and Wubs, Martijn and Holleitner, Alexander W. and Huck, Alexander and Kastl, Christoph and Stenger, Nicolas},
  journal = {Phys. Rev. Mater.},
  volume = {9},
  issue = {5},
  pages = {056203},
  numpages = {15},
  year = {2025},
  doi = {10.1103/PhysRevMaterials.9.056203},
  url = {https://link.aps.org/doi/10.1103/PhysRevMaterials.9.056203}
}

@article{https://doi.org/10.1002/smll.202301926,
author = {Bui, Thuy An and Leuthner, Gregor T. and Madsen, Jacob and Monazam, Mohammad R. A. and Chirita, Alexandru I. and Postl, Andreas and Mangler, Clemens and Kotakoski, Jani and Susi, Toma},
title = {Creation of Single Vacancies in hBN with Electron Irradiation},
journal = {Small},
volume = {19},
number = {39},
pages = {2301926},
doi = {https://doi.org/10.1002/smll.202301926},
url = {https://onlinelibrary.wiley.com/doi/abs/10.1002/smll.202301926},
eprint = {https://onlinelibrary.wiley.com/doi/pdf/10.1002/smll.202301926},
year = {2023}
}

\end{document}